\documentclass[journal=aelccp,manuscript=letter]{achemso}

\usepackage{chemformula} 
\usepackage[T1]{fontenc} 
\usepackage{tabularx}

\SectionsOn

\newcolumntype{Y}{>{\raggedleft\arraybackslash}X}

\author{Michael W. Swift}
\affiliation{Center for Computational Materials Science, Naval Research Laboratory, Washington, DC 20375, USA.}
\email{michael.swift@nrl.navy.mil}
\author{John L. Lyons}
\affiliation{Center for Computational Materials Science, Naval Research Laboratory, Washington, DC 20375, USA.}
\email{john.lyons@nrl.navy.mil}

\title[Lone pairs in halide perovskites]
  {Lone-Pair Stereochemistry Induces Ferroelectric Distortion and the Rashba Effect in  Inorganic Halide Perovskites}
  
\begin{document}







\begin{abstract}
  The lone-pair s states of germanium, tin, and lead underlie many of the unconventional properties of the inorganic metal halide perovskites.  Dynamic stereochemical expression of the lone pairs is well established for perovskites based on all three metals, but previously only the germanium perovskites were thought to express the lone pair crystallographically.  In this work, we use advanced first-principles calculations with a hybrid functional and spin--orbit coupling to predict stable monoclinic polar phases of \ch{CsSnI3} and \ch{CsSnBr3}, which exhibit a ferroelectric distortion driven by stereochemical expression of the tin lone pair.  We also predict similar metastable ferroelectric phases of \ch{CsPbI3} and \ch{CsPbBr3}.  In addition to ferroelectricity, these phases exhibit the Rashba effect. Spin splitting in both the conduction and valence bands suggests that nanostructures based on these phases could host bright ground-state excitons.  Finally, we discuss paths toward experimental realization of these phases \emph{via} electric fields and tensile strain.

\centering
\clearpage
TOC Graphic

\includegraphics{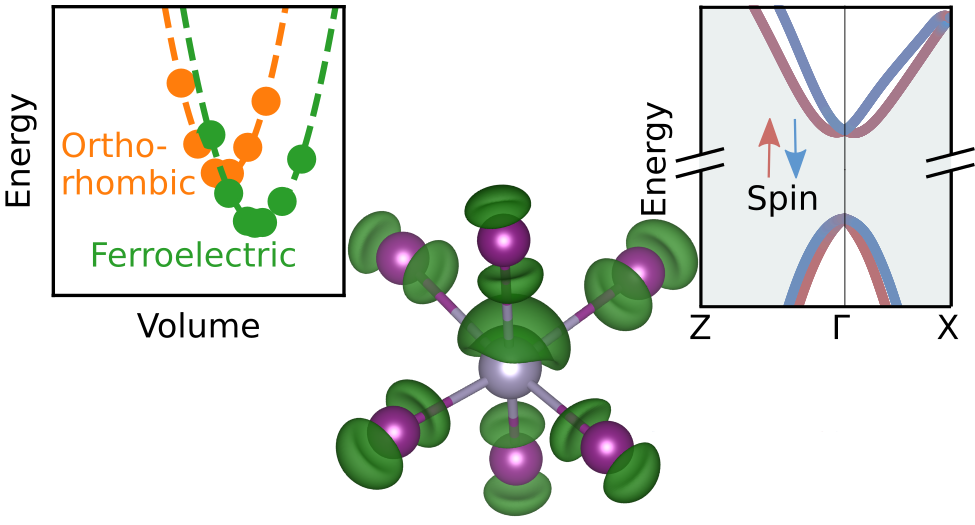}

\end{abstract}

\begin{figure}
\includegraphics[width=\textwidth]{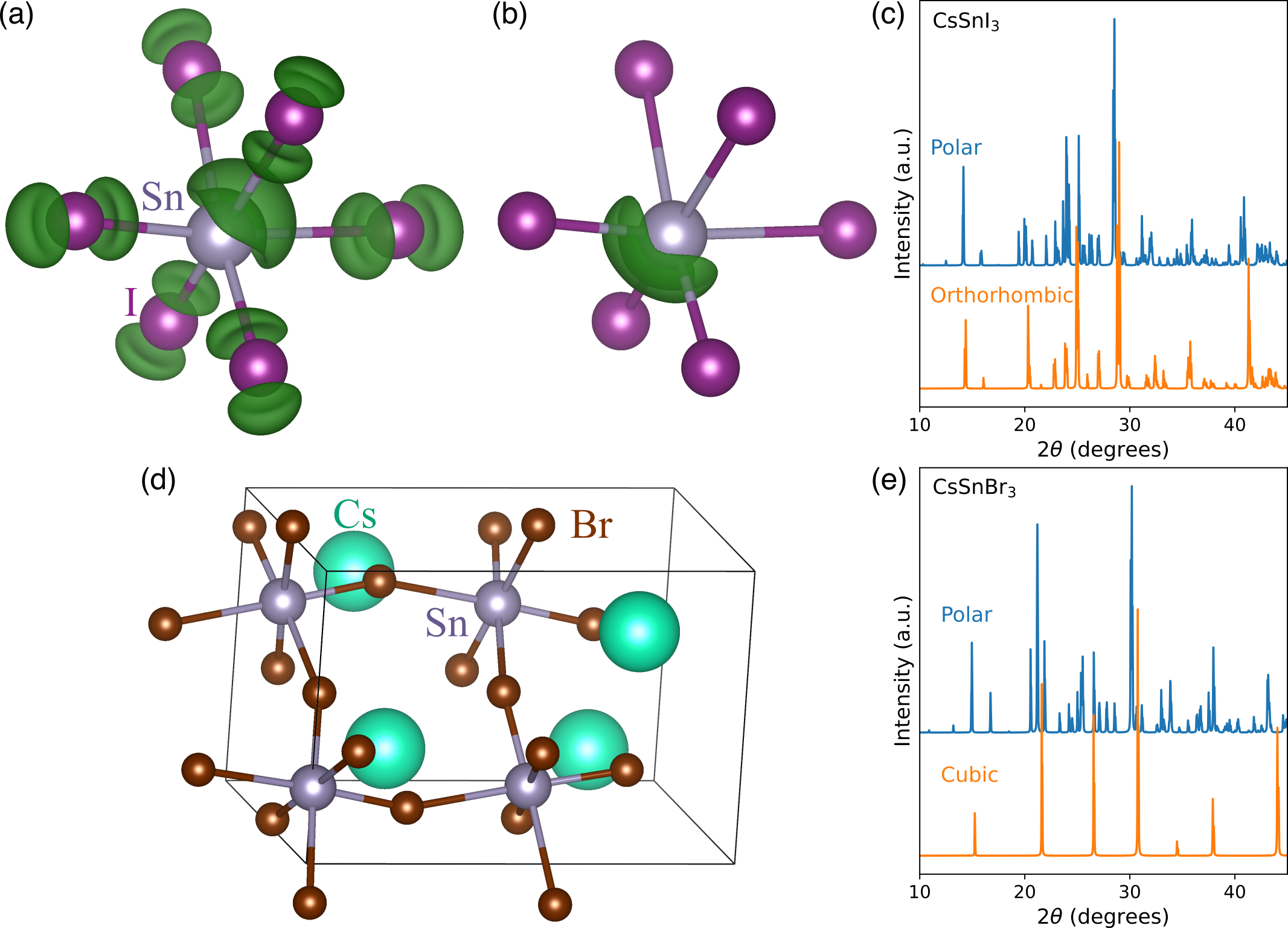}
\caption{(a-b) Lone-pair electronic states in the predicted ferroelectric phase of \ch{CsSnI3}, visualized as constant-density isosurfaces (green).  Tin are shown as gray spheres and iodine as purple spheres.   Panel (a) shows the valence-band maximum states, exhibiting I 5p character hybridized with a stereochemically expressed Sn 5s lone pair.  The underlying SnI$_6$ octahedron shows a ferroelectric distortion along the lone-pair direction.  The Sn--I bonds in the direction of the valence lone pair are elongated to $\sim$3.5 \AA.  Panel (b) shows the deep Sn 5s states, the bonding state of the stereochemically expressed lone pair.  The Sn--I bonds in the direction of this deep lone pair are shortened to $\sim$2.9 \AA.  (c) Simulated powder XRD spectra for \ch{CsSnI3}, comparing the known orthorhombic and the predicted ferroelectric phases.  (d) Full unit cell of ferroelectric \ch{CsSnBr3}, showing the modified octahedral tilting that accompanies the ferroelectric distortion.   (e) Simulated powder XRD spectra for \ch{CsSnBr3}, comparing the known cubic and the predicted ferroelectric phases.}
\label{fig:visualization}
\end{figure}

Halide perovskites are a versatile class of optoelectronic materials whose potential applications vary with composition and whose performance often depends on crystal structure.  Hybrid organic-inorganic lead halide perovskites---$ABX_3$ with methylammonium or formamidinium on the $A$ site, lead on the $B$ site, and a halide on the $X$ site---are well known solar-cell materials with rapidly growing efficiencies.~\cite{nrel,kojima_organometal_2009,chung_all-solid-state_2012,green_perovskite_2017} The most efficient halide perovskites solar cells often contain mixed compositions in order to maintain phase stability.\cite{green_perovskite_2017} All-inorganic lead halide perovskites (with cesium on the $A$ site) have attracted attention as light emitters, especially for colloidal nanostructures~\cite{LoredanaProtesescu:2015et,Kovalenko:2017ge,Becker:2018dg,Utzat:2019,swift_dark_2022}.   The lead-free tin halide perovskites are promising and well studied, but have received less attention than their lead-based cousins~\cite{chung_cssni_2012,gupta_cssnbr_2016,fabini_dynamic_2016}.  Both the tin and lead perovskites show complex structure-property relationships that are key to their exceptional performance\cite{Sercel:2019je,ke_unleaded_2019,monacelli_first-principles_2023}. In addition to their technological relevance, the halide perovskites have attracted significant fundamental interest due to their unconventional lattice dynamics.~\cite{Becker:2018dg,fabini_underappreciated_2020,lanigan-atkins_two-dimensional_2021,schilcher_significance_2021}

Many of the unusual properties of halide perovskites are due to the ``lone-pair'' s states of the group IV element on the $B$ site~\cite{fabini_dynamic_2016,radha_distortion_2018,fabini_underappreciated_2020,gao_metal_2021,fu_stereochemical_2021}. The lone-pair states hybridize with the halide p states, and the resulting antibonding state forms the valence-band maximum (VBM).  The bonding state is deep in the valence band---8 to 10 eV below the VBM---and has almost entirely s character.  Like the lone pairs that lead to the bent geometry of the water molecule or the
pyramidal geometry of ammonia, these group-IV s states can be stereochemically expressed: by localizing on one side of the atom, they distort the octahedra and break structural inversion symmetry (see Figure~\ref{fig:visualization}).  This distortion requires extra space, setting up a competition between lone-pair expression and lattice strain. Understanding how these lone pairs manifest in the halide perovskites, and whether they might lead to emergent behavior such as ferroelectricity or the Rashba effect, is the subject of intense debate.\cite{marronnier_anharmonicity_2018,marronnier_influence_2019,steele_role_2019,mohd_yusoff_observation_2021,fu_stereochemical_2021,ambrosio_ferroelectricferroelastic_2022}

In this work, we predict stable ferroelectric phases of \ch{CsSnI3} and \ch{CsSnBr3}, in which the lone pairs are crystallographically expressed, much like in the CsGe$X_3$ phases.  We also predict metastable ferroelectric phases of \ch{CsPbI3} and \ch{CsPbBr3} that may be stabilized by tensile strain.  Accurate results require the use of a hybrid functional; we find that semilocal functionals have underestimated the tendency towards stereochemical expression of tin and lead lone pairs.  Because these heavier group-IV elements have strong spin--orbit coupling, the polar environment leads to Rashba-type spin splitting in both the valence and conduction bands.  This ``double-Rashba'' behavior may be the key to generating a bright ground-state exciton in nanocrystals~\cite{Becker:2018dg,Sercel:2019je,Sercel:2019hs,swift_rashba_2021,swift_BGE_2023}, and could explain evidence of a Rashba effect in \ch{CsPbBr3} nanocrystals despite their crystallographic inversion symmetry~\cite{Becker:2018dg,Isarov:2017do}.  The low-symmetry polar phases also exhibit ferroelectricity, perhaps explaining previously puzzling signatures of ferroelectricity~\cite{li_evidence_2020} and permanent dipole moments~\cite{lv_probing_2021} in these materials.

For CsGe$X_3$, the germanium 4s states are relatively small, so there is ample room for the lone pair to be stereochemically expressed.  In the ground state, the lone pair is crystallographically expressed (i.e. frozen in the same direction on all germanium sites), resulting in a polar space group~\cite{liu_hybrid_2022,stoumpos_hybrid_2015}.  For CsSn$X_3$ and CsPb$X_3$, the tin 5s and lead 6s states are larger than the germanium 4s. Previous studies have found that the lattice does not provide sufficient room for coordinated off-centering and therefore the overall space group is centrosymmetric.  However, there is evidence that the lone pair may still be expressed in the form of local, dynamical polar fluctuations that are averaged out in x-ray diffraction measurements, but which nevertheless have dramatic impacts on the structural and electronic properties of the halide perovskites.~\cite{fabini_dynamic_2016,marronnier_anharmonicity_2018,marronnier_influence_2019,steele_role_2019,fabini_underappreciated_2020,fu_stereochemical_2021,gao_metal_2021,mohd_yusoff_observation_2021}

Previous computational studies of lone pairs in the halide perovskites using DFT have mostly employed semilocal functionals such as PBE~\cite{PBE}.  It is well established that, while PBE predicts band gaps that are close to correct in the lead and tin halide perovskites due to a coincidental cancellation of errors, the absolute band-edge energies are incorrect when using PBE alone~\cite{Du:2015hs,DeAngelis:2018gv}.  A combination of a hybrid functional and the inclusion of spin--orbit coupling (which we refer to as HSE+SOC) overcame this problem in studies of charged point defects, where the relative position of band edges and defect charge-state transition levels is key to their behavior~\cite{Du:2015hs,DeAngelis:2018gv,zhang_iodine_2023,lyons_trends_2023}.  In addition to underestimating the band gap, semilocal functionals are known to underestimate the localization of electronic states, whereas HSE describes localized states more accurately~\cite{henderson_accurate_2011}.  This leads naturally to the question: how does an improved functional change the description of lone pairs in these materials?

Using HSE+SOC we have identified ground-state polar ferroelectric phases of \ch{CsSnI3} and \ch{CsSnBr3} with the monoclinic space group $Pc$ (\#7).  The unit cell is similar to the orthorhombic phase, but with tin off-centering and inversion asymmetry driven by coordinated stereochemical expression of the tin 5s lone pairs.  In \ch{CsSnI3}, the ferroelectric phase is lower in energy than the orthorhombic phase by 34.6 meV per formula unit, and lower than the non-perovskite ``yellow'' phase by 7.1 meV per formula unit.  In \ch{CsSnBr3}, the ferroelectric phase is lower in energy than the cubic phase by 63.0 meV per formula unit, and lower in energy than the orthorhombic phase by 66.0 meV per formula unit.  Relaxed structures for these predicted phases are included in the supporting information.

Figure~\ref{fig:visualization} shows a visualization of the lone pair states and a distorted SnI$_6$ octahedron in \ch{CsSnI3}.  Panel (a) shows the charge density of the two highest occupied states, which make up the valence-band maximum (VBM).  The tin 5s lone pair is seen by the asymmetric lobe of charge density on the tin atom.  The tin 5s hybridizes with iodine 5p, and the antibonding states make up the VBM.  This is consistent with the observation that the Sn--I bonds in the direction of the antibonding state of the lone pair are elongated, with lengths 3.50, 3.49, and 3.45 \AA.  Panel (b) shows the charge density of the deep tin 5s states, which are approximately 8 eV below the VBM in energy.  These states are almost purely tin 5s in character.  The Sn--I bonds in this direction are shortened, with lengths 2.94, 2.94, and 2.93 \AA.

\begin{figure}
\includegraphics[width=0.5\textwidth]{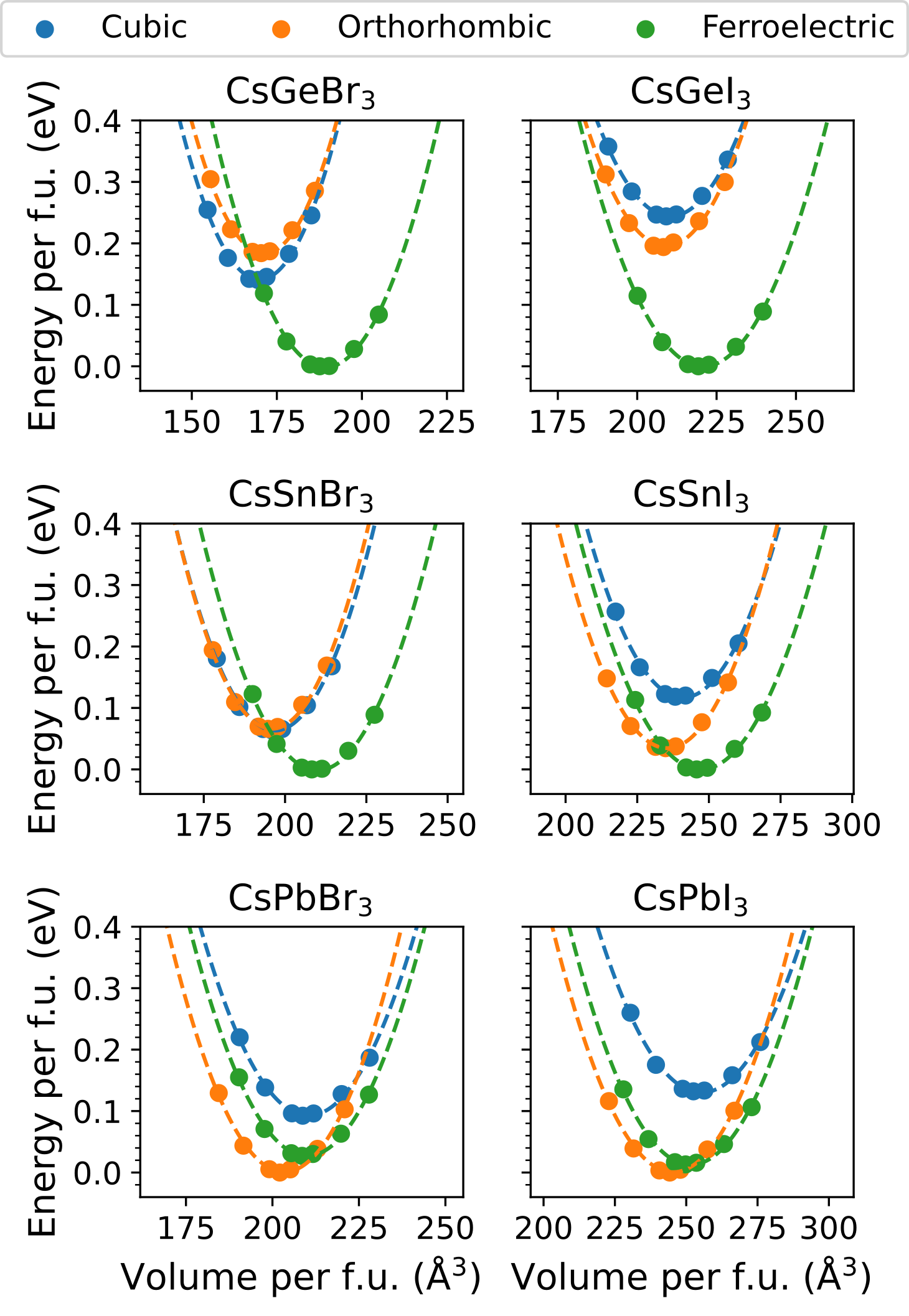}
\caption{Energy per formula unit of cubic, orthorhombic, and ferroelectric phases of Cs$BX_3$, plotted as a function of volume per formula unit, with dashed lines are parabolic fits.  We find the ferroelectric phase with the stereochemically expressed lone pair is the most stable for both the germanium and tin materials.  The ferroelectric phase is predicted to be metastable in the lead materials, but becomes energetically favored upon tensile strain.  The critical strain, defined as the strain at which the polar and non-polar phases are equal in energy, is given in Table~\ref{tab:critical_strain} for each material. }
\label{fig:strain}
\end{figure}

Stereochemical expression of the lone pair requires more space than if the s state remains centrosymmetric.  This can be observed in Figure~\ref{fig:strain}, which plots the energy of the cubic, orthorhombic, and ferroelectric phases as a function of volume.   In each case, the centrosymmetric phases have a lower volume than the ferroelectric phase.  Application of tensile strain can therefore be expected to stabilize the ferroelectric phase.  Table~\ref{tab:critical_strain} shows the critical strain $\varepsilon_\text{crit}$ for the stereochemical lone-pair expression, relative to the lowest-energy phase, which is defined as
\begin{equation}
    \varepsilon_\text{crit} = \left(\frac{V_1 - V_0}{V_0}\right)^{1/3}~,
\end{equation}
where $V_0$ is the equilibrium volume of the base phase and $V_1$ is the volume at which the phases have equal energy.  For Ge and Sn, the lowest-energy phase is ferroelectric, so negative $\varepsilon_\text{crit}$ values indicate the amount of compressive strain which, if applied to the ferroelectric phase, would stabilize the centrosymmetric phase.  Conversely, for Pb, the lowest-energy phase is orthorhombic, so positive $\varepsilon_\text{crit}$ values indicate the amount of tensile strain which, if applied to the orthorhombic phase,  would stabilize the ferroelectric phase.  Comparable figures using PBE (shown in Supporting Figure S1) predict the lowest-energy phases of the tin perovskites to be centrosymmetric, showing that semilocal functionals overestimate the energy of the ferroelectric phase.

\begin{table}
\caption{Critical strain for switching the lone-pair expression on or off in Cs$BX_3$, with $B$ = Ge, Sn, and Pb in columns and $X$ = Br, I in rows.  If the critical strain is less than 0 ($B$ = Ge or Sn), the ferroelectric phase is more stable than the orthorhombic phase, but the orthorhombic phase becomes more stable at the critical value of compressive strain.  If the critical strain is greater than 0 ($B$ = Pb), the orthorhombic phase is more stable, but the ferroelectric phase becomes more stable at the critical value of tensile train. }
\begin{tabularx}{\columnwidth}{l|YYY}\hline\hline
    & Ge    & Sn    & Pb    \\\hline
Br  & $-3.6\%$ & $-2.4\%$ & $+1.4\%$ \\
I   & $-8.3\%$ & $-1.8\%$ & $+1.0\%$ \\\hline\hline
\end{tabularx}
\label{tab:critical_strain}
\end{table}

\begin{figure*}
\includegraphics[width=\textwidth]{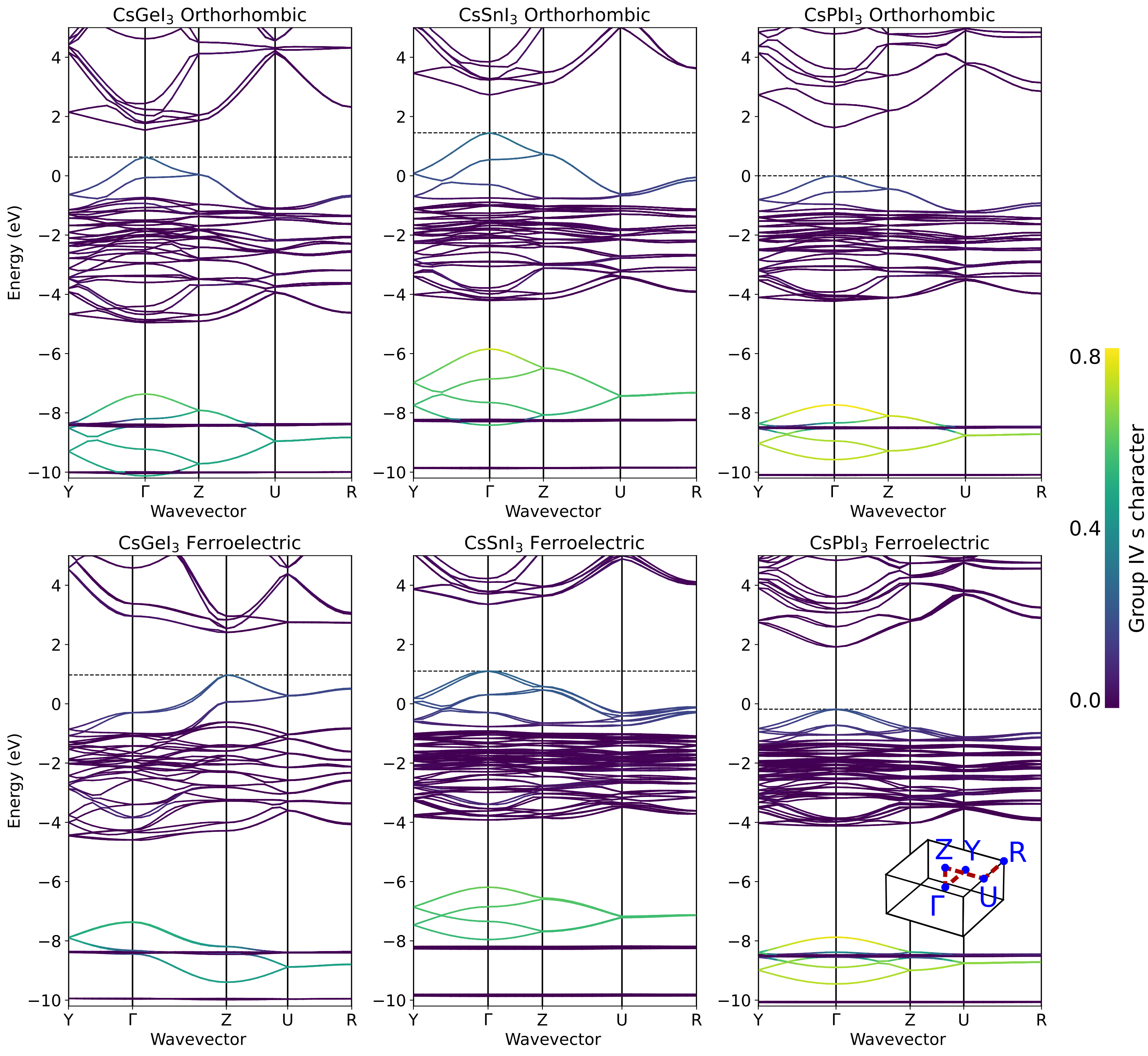}
\caption{Band structures of Cs$B$I$_3$ comparing orthorhombic and ferroelectric phases.  The color of the bands shows the fraction of the charge density that projects onto the s spherical harmonic of the group-IV element at the $B$ site, which occurs in the VBM and in deep states from $-6$ to $-10$ eV (see Figure~\ref{fig:visualization}).  Band structures are aligned to the deep Cs 5s states, which are set to $-23.6$ eV so that the VBM of \ch{CsPbI3} is at zero energy.  The valence-band states are primarily iodine 5p, the conduction bands are primarily made of the group-IV p states, and the flat bands near $-8$ and $-10$ eV are the Cs 5p states.  The first Brillouin zone of ferroelectric \ch{CsPbI3} is shown as an inset with labeled high-symmetry points.  }
\label{fig:BS}
\end{figure*}

The effect of the lone pairs may also be seen in the band structure, plotted for the iodides in Figure~\ref{fig:BS} and for the bromides in Supporting Figure S2.  The ferroelectric distortion increases the band gap, primarily by narrowing the bandwidths.  This is particularly pronounced in the deep group-IV s states, which narrow substantially going from the spherically symmetric state in the orthorhombic phase to the more localized, stereochemically expressed state in the ferroelectric phase. Since the energy difference between phases is only tens of meV (nearly invisible on this scale), the occupied bands' average energy only changes slightly between the two structures.

To investigate the ferroelectric switching behavior of the CsSn$X_3$ phases, we begin by finding the transition state for the switching process.  We start with the midpoint between the polar phase and its mirror image, then relax \emph{via} a single-image nudged-elastic-band calculation~\cite{h_jonsson_nudged_1998} with the polar phase and its mirror image as endpoints.  The result is the transition state.  The switching barrier per formula unit is 0.08 eV for \ch{CsSnI3} and 0.13 eV for \ch{CsSnBr3}.

The transition state is centrosymmetric, so it can also be used as a reference nonpolar state to calculate the spontaneous polarization of the ferroelectric phase using the modern theory of polarization~\cite{King-Smith_Vanderbilt_Polarization_1993} (see Supporting Figure S3 for details).  The spontaneous polarization is 14.3 $\mu$C/cm$^2$ for \ch{CsSnI3} and 17.3 $\mu$C/cm$^2$ for \ch{CsSnBr3}.  These values are significantly larger than the previous estimate of 4.4 $\mu$C/cm$^2$ for ferroelectric polarization in CH$_3$NH$_3$PbI$_3$~\cite{stroppa_ferroelectric_2015}, and are closer to the polarization observed in traditional ferroelectric oxides such as BaTiO$_3$ (27 $\mu$C/cm$^2$)\cite{Wieder_1955}. A rough estimate for the intrinsic coercive field is $\mathcal{E}_c \approx E_B/(\Omega P_s)$, where $E_B$ is the barrier energy, $\Omega$ is the volume, and $P_s$ is the spontaneous polarization.  This gives 3.8 MV/cm for \ch{CsSnI3} and 5.9 MV/cm for \ch{CsSnBr3}.  Note that these values are upper bounds for the coercive field; experimental values may be significantly smaller and will depend strongly on temperature.~\cite{ducharme_intrinsic_2000}

Whether ferroelectricity is the origin of the exceptional photovoltaic performance in the halide perovskites has been the subject of intense debate.\cite{stroppa_ferroelectric_2015,fu_stereochemical_2021,ambrosio_ferroelectricferroelastic_2022}. Prior studies have concluded that, due to their cubic crystal phase\cite{ambrosio_ferroelectricferroelastic_2022} and insufficient lattice instability\cite{fu_stereochemical_2021}, ferroelectricity was irrelevant for explaining the performance of halide perovskites. These conclusions may need to be reassessed, at least for the Sn-based perovskites in light of the new phases discovered here for \ch{CsSnBr3} and \ch{CsSnI3}.

\begin{figure}
\includegraphics[width=0.4\textwidth]{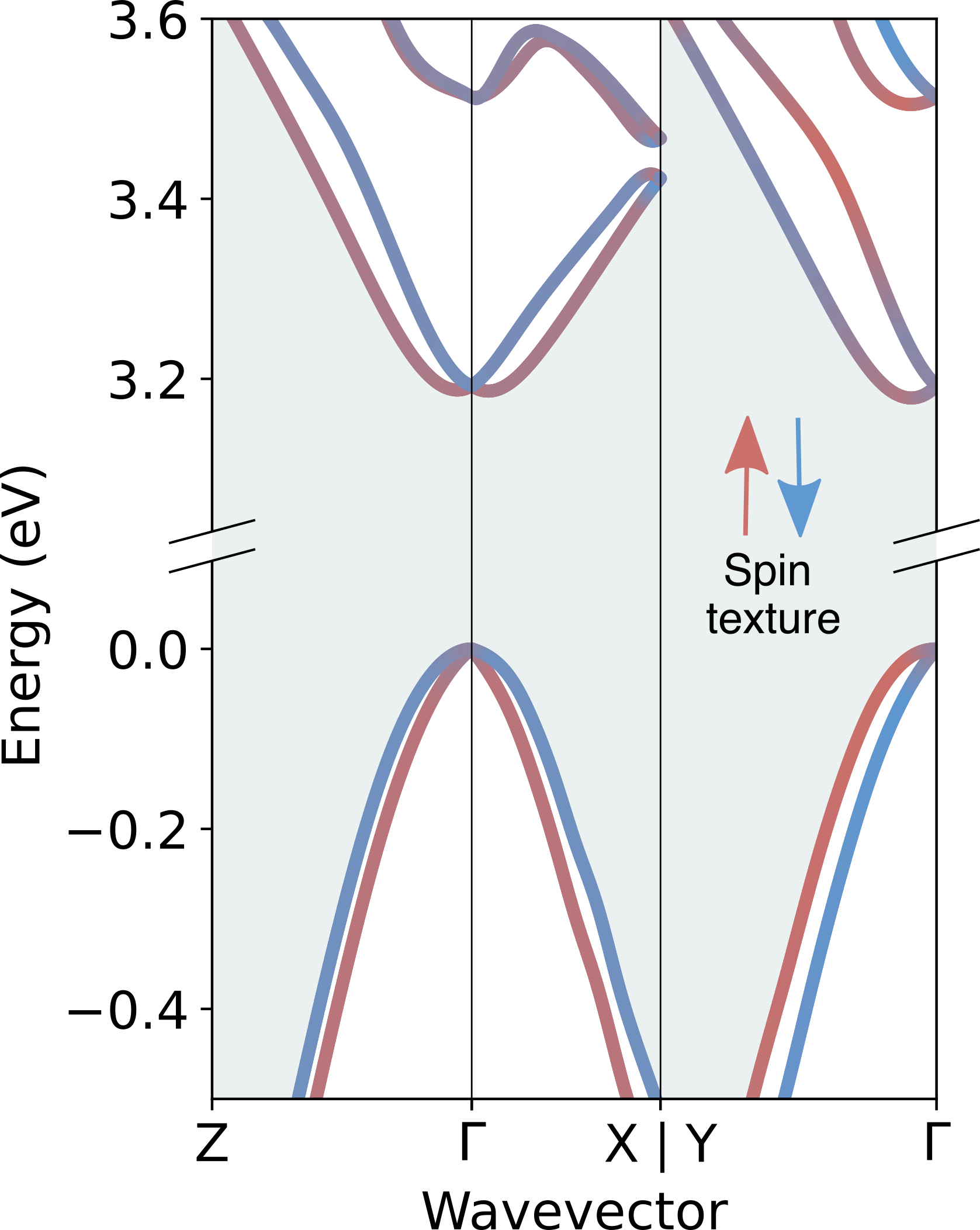}
\caption{Band structure detail of \ch{CsSnBr3}, focusing on the CBM and VBM states.  The characteristic Rashba ``double well'' is clear in both the CBM and VBM.  The color of the bands is determined by the spin projection in the $x$ direction.  The spin texture is contra-helical in the two bands. }
\label{fig:Rashba_BS}
\end{figure}

A full assessment of the impact of the distorted phase on photovoltaic performance is beyond the scope of this work, but it is clear that the polar distortion can lead to emergent behavior such as the Rashba effect, which can in turn have important implications on optoelectronic properties~\cite{marronnier_anharmonicity_2018,marronnier_influence_2019,steele_role_2019,mohd_yusoff_observation_2021}. When the lone pair is stereochemically expressed, the group-IV atom on the $B$ site experiences an effective electric field because of the polar environment.  For the heavier group-IV elements this leads to Rashba-type spin--orbit splitting.  Letting $z$ be the direction of the inversion asymmetry, the Rashba interaction for electrons (and holes) can be expressed through the Hamiltonian
\begin{equation}
    \hat H_R = \alpha_R (\hat\sigma_x \hat p_y - \hat \sigma_y \hat p_x)~.
\end{equation}
The Rashba energy $E_R$ is related to the Rashba coefficient $\alpha_R$ through $E_R = \alpha_R^2 m / 2\hbar^2$, where $m$ is the effective mass.  While germanium is too light to experience noticeable spin--orbit coupling, ferroelectric CsSn$X_3$ and CsPb$X_3$ exhibit detectable Rashba splitting in both the conduction and valence bands.  This ``double-Rashba'' character is indicative of a possible bright ground-state exciton that could be achieved upon appropriate nanostructuring~\cite{swift_BGE_2023}.  The effective mass of the electron and hole, as well as their Rashba energies, are given in Table~\ref{tab:BS_Rashba_params}, a detailed view of the \ch{CsSnBr3} band structure color-coded by spin is shown in Figure~\ref{fig:Rashba_BS}, and spin texture plots may be found in Supporting Figure S4.  The spin textures show mixed Rashba-Dresselhaus spin splitting, with spin-splitting along the direction of the polarization in the valence band but canted in the conduction band.

\begin{table}
\caption{Band structure parameters for ferroelectric phases: band gap $E_g$, electron effective mass $m_e$, hole effective mass $m_h$, electron Rashba energy $E_R^e$, and hole Rashba energy $E_R^h$. }
\begin{tabularx}{\columnwidth}{l|YYYYY}
\hline\hline
& $E_g$ (eV) & $m_e$ $(m_0)$ & $m_h$ $(m_0)$ & $E_R^e$ (meV) & $E_R^h$ (meV) \\
\hline
\ch{CsSnI3} & 2.25 & 0.267 & 0.213 & 0.93 & 2.02\\
\ch{CsSnBr3} & 3.18 & 0.360 & 0.476 & 13.33 & 5.94 \\
\ch{CsPbI3} & 2.02 & 0.243 & 0.327 & 2.30 & 5.38 \\
\ch{CsPbBr3} & 2.68 & 0.313 & 0.410 & 0.26 & 7.14 \\
\hline\hline
\end{tabularx}
\label{tab:BS_Rashba_params}
\end{table}

In Ref.~\citenum{Becker:2018dg}, Rashba coefficients of 0.38 eV$\cdot$\AA~were assumed for the conduction and valence band, though a mechanism for inversion asymmetry was not proposed.  The strength of the Rashba coupling in our predicted metastable polar \ch{CsPbBr3} (see Table~\ref{tab:BS_Rashba_params}) corresponds to Rashba coefficients of 0.11 eV$\cdot$\AA~in the conduction band and 0.52 eV$\cdot$\AA~in the valence band, suggesting that this metastable phase could be a relevant source of inversion asymmetry in \ch{CsPbBr3} nanocrystals.  Perovskite nanocrystals are known to be under tensile strain from the capping ligands~\cite{swarnkar_quantum_2016,zhao_size-dependent_2020,jia_inhibiting_2022}, lending additional credence to this possibility.  Furthermore, due to the symmetry properties of the perovskite Bloch functions, contra-helical spin textures imply co-helical angular momentum textures, and therefore equal signs of the electron and hole Rashba coefficient~\cite{Sercel:2019je,swift_rashba_2021}.  This is also consistent with the Rashba splitting assumed in Ref.~\citenum{Becker:2018dg}.

The ferroelectric phase of \ch{CsSnBr3} may have already been found experimentally; a transition to an unknown monoclinic phase has been reported below $-26^\circ$ C~\cite{mori_x-ray_1986}.  Identification of the ferroelectric phase of \ch{CsSnI3} may be more challenging because of competition from the non-perovskite ``yellow'' phase $\delta$-\ch{CsSnI3}.  Temperature-dependent calculations, including estimates of the Curie temperature of the ferroelectric phases, may help predict the conditions required to achieve these phases experimentally.
Simulated powder X-ray diffraction patterns for \ch{CsSnBr3} and \ch{CsSnI3} are provided in Figure~\ref{fig:visualization}c,e to aid in future experimental searches.  In the ferroelectric phases, the main peaks are shifted to smaller angles due to the larger lattice, and many smaller peaks appear due to the monoclinic symmetry.  
Applying an electric field during growth could provide a preferred direction for the lone pairs to align and further lower the energy of the polar phase.  Extended ferroelectric domains may also be more likely near the surface of nanocrystals, with the ligand shell providing tensile strain and the surface breaking inversion symmetry.  Epitaxial growth on a substrate with lattice parameters chosen to provide moderate tensile strain could also enhance the formation of the ferroelectric phase; the impact of biaxial strain on the lone pair stereochemistry should be a fruitful avenue for further research.

\section*{Computational Methods}

Calculations were performed in the Vienna ab initio simulation package (VASP)~\cite{VASP} using PAW pseudopotentials~\cite{PAW} and the HSE hybrid functional~\cite{HSE} with spin--orbit coupling included.  The plane-wave energy cutoff was 400 eV.  Unless otherwise noted, all relaxations continued until forces were less than 0.03 eV/\AA.  The switching barrier for the ferroelectric phases were found by interpolating between appropriately translated mirror images and performing a single-image nudged-elastic-band calculation~\cite{h_jonsson_nudged_1998} relaxed to achieve energy convergence of 0.01 eV.  Visualization and simulated powder XRD spectra were generated using VESTA~\cite{momma_vesta_2011}.  Symmetry analysis used spglib~\cite{togo_textttspglib_2018} in pymatgen~\cite{Ong2013}.  Effective masses are extracted from parabolic fits to the band structure, and Rashba energies are extracted from linear fits to the spin splitting as a function of $k$.

The HSE mixing and screening parameters were tuned for each $B$-site element to reproduce the experimental band gap in the room-temperature phases~\cite{Stoumpos:2013fz,marronnier_anharmonicity_2018,zhao_preparation_2019,fabini_dynamic_2016,gupta_cssnbr_2016,chung_cssni_2012,liu_hybrid_2022,stoumpos_hybrid_2015}; see Table~\ref{tab:params}.  Parameters were tuned for \ch{CsPbBr3}, \ch{CsSnI3}, and \ch{CsGeBr3}, and then the same parameters were used for \ch{CsPbI3}, \ch{CsSnBr3}, and \ch{CsGeI3} respectively.  Note that for \ch{CsPbI3}, ref.~\citenum{marronnier_anharmonicity_2018} reports lattice parameters in the $Pbnm$ space group, which we have recast to $Pnma$ to match our calculated structure.  For \ch{CsGeI3}, the initial structure was in the $R3m$ space group, but the relaxed structure was $Cm$.  Tests with ``standard'' HSE parameters (25\% mixing and 0.2 \AA$^{-1}$ screening) and the HSE parameters used in ref.~\citenum{zhang_iodine_2023} (55\% mixing and 0.2 \AA$^{-1}$ screening) also found the ferroelectric phases of CsSn$X_3$ to be the ground states, though the magnitude of the energy difference varied (see Supporting Table S1).

\begin{table}
\caption{HSE mixing and screening parameters used for each material, together with the resulting band gaps and lattice parameters. }
\begin{tabular}{|l|ll|ll|}\hline\hline
       & \multicolumn{2}{c|}{\ch{CsPbBr3} ($Pnma$)} & \multicolumn{2}{c|}{\ch{CsPbI3} ($Pnma$)}  \\
          & Theory         & Expt.~\cite{Stoumpos:2013fz}  & Theory   & Expt.~\cite{marronnier_anharmonicity_2018,zhao_preparation_2019}     \\\hline
Mixing    & 35\%           &             & 35\%           &                \\
Screening & 0.1 \AA$^{-1}$ &             & 0.1 \AA$^{-1}$ &                \\
Gap (eV)  & 2.32           & 2.25        & 1.80           & 1.75           \\
$a$ (\AA) & 8.4756         & 8.2440      & 9.0472         & 8.8518         \\
$b$ (\AA) & 11.768         & 11.7351     & 12.559         & 12.501         \\
$c$ (\AA) & 8.1061         & 8.1982      & 8.5983         & 8.6198         \\\hline
       & \multicolumn{2}{c|}{\ch{CsSnBr3} ($Pnma$)} & \multicolumn{2}{c|}{\ch{CsSnI3} ($Pnma$)}  \\
          & Theory         & Expt.~\cite{fabini_dynamic_2016,gupta_cssnbr_2016}  & Theory         & Expt.~\cite{chung_cssni_2012}     \\\hline
Mixing    & 41\%           &             & 41\%           &                \\
Screening & 0.1 \AA$^{-1}$ &             & 0.1 \AA$^{-1}$ &                \\
Gap (eV)  & 1.63           & 1.73        & 1.31           & 1.3            \\
$a$ (\AA) & 8.2529          & 8.1965      & 8.742          & 8.6885         \\
$b$ (\AA) & 11.639          & 11.583      & 12.394         & 12.3775        \\
$c$ (\AA) & 8.1103          & 8.0243      & 8.671          & 8.6384         \\\hline
       & \multicolumn{2}{c|}{\ch{CsGeBr3} ($R3m$) } & \multicolumn{2}{c|}{\ch{CsGeI3} ($Cm$)}  \\
          & Theory         & Expt.~\cite{liu_hybrid_2022}  & Theory   & Expt.~\cite{stoumpos_hybrid_2015}     \\\hline
Mixing    & 38\%           &             & 38\%           &                \\
Screening & 0.2 \AA$^{-1}$ &             & 0.2 \AA$^{-1}$ &                \\
Gap (eV)  & 2.49           & 2.42        & 1.47           & 1.60           \\
$a$ (\AA) & 7.9662         & 7.89870     & 8.8875         & 8.3582        \\
$b$ (\AA) & 7.9662         & 7.89870     & 8.5021         & 8.3582        \\
$c$ (\AA) & 10.244         & 9.98890     & 11.607         & 10.61         \\\hline\hline
\end{tabular}
\label{tab:params}
\end{table}

In Figure~\ref{fig:strain}, a fully relaxed calculation is the base of the parabola for each phase except the ferroelectric CsPb$X_3$.  Higher-energy points are from unrelaxed total-energy calculations of the base structure with hydrostatic strain applied.  For CsPb$X_3$, orthorhombic and cubic structures are fully relaxed, but the ferroelectric structures return to the orthorhombic phase when fully relaxed.  A metastable base structure for the ferroelectric phase was found by starting with the CsSn$X_3$ structure, adjusting the lattice parameters based on the ratio of the volumes of the orthorhombic phases, relaxing the atomic positions and cell shape at constant volume, and finally relaxing the cell volume with fixed relative atomic coordinates.  Since under tensile strain the ferroelectric structure is lower in energy than the orthorhombic structure, this procedure identifies a metastable ferroelectric phase.


\begin{acknowledgement}

This work was supported by the ONR/NRL 6.1 Base Research Program. The authors thank Dr.~Sasha Efros, Dr.~Peter Sercel, and Dr.~Noam Bernstein for helpful discussions.

\end{acknowledgement}

\begin{suppinfo}

The following files are available free of charge.
\begin{itemize}
  \item \texttt{Lone-pair\_SI.pdf}: Supporting Figures S1-S4 and Supporting Table S1
  \item \texttt{CsSnBr3\_Pc.cif}: Crystallographic information file (CIF) of the polar phase of \ch{CsSnBr3}
  \item \texttt{CsSnI3\_Pc.cif}: CIF of the polar phase of \ch{CsSnI3}
  \item \texttt{CsPbBr3\_Pc.cif}: CIF of the metastable polar phase of \ch{CsPbBr3}
  \item \texttt{CsPbI3\_Pc.cif}: CIF of the metastable polar phase of \ch{CsPbI3}
\end{itemize}

\end{suppinfo}

\providecommand{\latin}[1]{#1}
\makeatletter
\providecommand{\doi}
  {\begingroup\let\do\@makeother\dospecials
  \catcode`\{=1 \catcode`\}=2 \doi@aux}
\providecommand{\doi@aux}[1]{\endgroup\texttt{#1}}
\makeatother
\providecommand*\mcitethebibliography{\thebibliography}
\csname @ifundefined\endcsname{endmcitethebibliography}
  {\let\endmcitethebibliography\endthebibliography}{}


\begin{mcitethebibliography}{59}
\providecommand*\natexlab[1]{#1}
\providecommand*\mciteSetBstSublistMode[1]{}
\providecommand*\mciteSetBstMaxWidthForm[2]{}
\providecommand*\mciteBstWouldAddEndPuncttrue
  {\def\EndOfBibitem{\unskip.}}
\providecommand*\mciteBstWouldAddEndPunctfalse
  {\let\EndOfBibitem\relax}
\providecommand*\mciteSetBstMidEndSepPunct[3]{}
\providecommand*\mciteSetBstSublistLabelBeginEnd[3]{}
\providecommand*\EndOfBibitem{}
\mciteSetBstSublistMode{f}
\mciteSetBstMaxWidthForm{subitem}{(\alph{mcitesubitemcount})}
\mciteSetBstSublistLabelBeginEnd
  {\mcitemaxwidthsubitemform\space}
  {\relax}
  {\relax}

\bibitem[nre()]{nrel}
Best Research-Cell Efficiency Chart.
  \url{https://www.nrel.gov/pv/cell-efficiency.html}, {National} Renewable
  Energy Laboratory (NREL)\relax
\mciteBstWouldAddEndPuncttrue
\mciteSetBstMidEndSepPunct{\mcitedefaultmidpunct}
{\mcitedefaultendpunct}{\mcitedefaultseppunct}\relax
\EndOfBibitem
\bibitem[Kojima \latin{et~al.}(2009)Kojima, Teshima, Shirai, and
  Miyasaka]{kojima_organometal_2009}
Kojima,~A.; Teshima,~K.; Shirai,~Y.; Miyasaka,~T. Organometal {Halide}
  {Perovskites} as {Visible}-{Light} {Sensitizers} for {Photovoltaic} {Cells}.
  \emph{J. Am. Chem. Soc.} \textbf{2009}, \emph{131}, 6050--6051\relax
\mciteBstWouldAddEndPuncttrue
\mciteSetBstMidEndSepPunct{\mcitedefaultmidpunct}
{\mcitedefaultendpunct}{\mcitedefaultseppunct}\relax
\EndOfBibitem
\bibitem[Chung \latin{et~al.}(2012)Chung, Lee, He, Chang, and
  Kanatzidis]{chung_all-solid-state_2012}
Chung,~I.; Lee,~B.; He,~J.; Chang,~R. P.~H.; Kanatzidis,~M.~G. All-solid-state
  dye-sensitized solar cells with high efficiency. \emph{Nature} \textbf{2012},
  \emph{485}, 486--489\relax
\mciteBstWouldAddEndPuncttrue
\mciteSetBstMidEndSepPunct{\mcitedefaultmidpunct}
{\mcitedefaultendpunct}{\mcitedefaultseppunct}\relax
\EndOfBibitem
\bibitem[Green and Ho-Baillie(2017)Green, and
  Ho-Baillie]{green_perovskite_2017}
Green,~M.~A.; Ho-Baillie,~A. Perovskite {Solar} {Cells}: {The} {Birth} of a
  {New} {Era} in {Photovoltaics}. \emph{ACS Energy Lett.} \textbf{2017},
  \emph{2}, 822--830\relax
\mciteBstWouldAddEndPuncttrue
\mciteSetBstMidEndSepPunct{\mcitedefaultmidpunct}
{\mcitedefaultendpunct}{\mcitedefaultseppunct}\relax
\EndOfBibitem
\bibitem[Protesescu \latin{et~al.}(2015)Protesescu, Yakunin, Bodnarchuk, Krieg,
  Caputo, Hendon, Yang, Walsh, and Kovalenko]{LoredanaProtesescu:2015et}
Protesescu,~L.; Yakunin,~S.; Bodnarchuk,~M.~I.; Krieg,~F.; Caputo,~R.;
  Hendon,~C.~H.; Yang,~R.~X.; Walsh,~A.; Kovalenko,~M.~V. {Nanocrystals of
  Cesium Lead Halide Perovskites (CsPbX 3, X = Cl, Br, and I): Novel
  Optoelectronic Materials Showing Bright Emission with Wide Color Gamut}.
  \emph{Nano Lett.} \textbf{2015}, \emph{15}, 3692--3696\relax
\mciteBstWouldAddEndPuncttrue
\mciteSetBstMidEndSepPunct{\mcitedefaultmidpunct}
{\mcitedefaultendpunct}{\mcitedefaultseppunct}\relax
\EndOfBibitem
\bibitem[Kovalenko \latin{et~al.}(2017)Kovalenko, Protesescu, and
  Bodnarchuk]{Kovalenko:2017ge}
Kovalenko,~M.~V.; Protesescu,~L.; Bodnarchuk,~M.~I. {Properties and potential
  optoelectronic applications of lead halide perovskite nanocrystals}.
  \emph{Science} \textbf{2017}, \emph{358}, 745--750\relax
\mciteBstWouldAddEndPuncttrue
\mciteSetBstMidEndSepPunct{\mcitedefaultmidpunct}
{\mcitedefaultendpunct}{\mcitedefaultseppunct}\relax
\EndOfBibitem
\bibitem[Becker \latin{et~al.}(2018)Becker, Vaxenburg, Nedelcu, Sercel,
  Shabaev, Mehl, Michopoulos, Lambrakos, Bernstein, Lyons, St{\"o}ferle, Mahrt,
  Kovalenko, Norris, Rain{\`o}, and Efros]{Becker:2018dg}
Becker,~M.~A. \latin{et~al.}  {Bright triplet excitons in caesium lead halide
  perovskites}. \emph{Nature} \textbf{2018}, \emph{553}, 189--193\relax
\mciteBstWouldAddEndPuncttrue
\mciteSetBstMidEndSepPunct{\mcitedefaultmidpunct}
{\mcitedefaultendpunct}{\mcitedefaultseppunct}\relax
\EndOfBibitem
\bibitem[Utzat \latin{et~al.}(2019)Utzat, Sun, Kaplan, Krieg, Ginterseder,
  Spokoyny, Klein, Shulenberger, Perkinson, Kovalenko, and Bawendi]{Utzat:2019}
Utzat,~H.; Sun,~W.; Kaplan,~A. E.~K.; Krieg,~F.; Ginterseder,~M.; Spokoyny,~B.;
  Klein,~N.~D.; Shulenberger,~K.~E.; Perkinson,~C.~F.; Kovalenko,~M.~V.;
  Bawendi,~M.~G. Coherent single-photon emission from colloidal lead halide
  perovskite quantum dots. \emph{Science} \textbf{2019}, \emph{363},
  1068--1072\relax
\mciteBstWouldAddEndPuncttrue
\mciteSetBstMidEndSepPunct{\mcitedefaultmidpunct}
{\mcitedefaultendpunct}{\mcitedefaultseppunct}\relax
\EndOfBibitem
\bibitem[Gramlich \latin{et~al.}(2022)Gramlich, Swift, Lampe, Lyons,
  Döblinger, Efros, Sercel, and Urban]{swift_dark_2022}
Gramlich,~M.; Swift,~M.~W.; Lampe,~C.; Lyons,~J.~L.; Döblinger,~M.;
  Efros,~Al.~L.; Sercel,~P.~C.; Urban,~A.~S. Dark and {Bright} {Excitons} in
  {Halide} {Perovskite} {Nanoplatelets}. \emph{Advanced Science} \textbf{2022},
  \emph{9}, 2103013\relax
\mciteBstWouldAddEndPuncttrue
\mciteSetBstMidEndSepPunct{\mcitedefaultmidpunct}
{\mcitedefaultendpunct}{\mcitedefaultseppunct}\relax
\EndOfBibitem
\bibitem[Chung \latin{et~al.}(2012)Chung, Song, Im, Androulakis, Malliakas, Li,
  Freeman, Kenney, and Kanatzidis]{chung_cssni_2012}
Chung,~I.; Song,~J.-H.; Im,~J.; Androulakis,~J.; Malliakas,~C.~D.; Li,~H.;
  Freeman,~A.~J.; Kenney,~J.~T.; Kanatzidis,~M.~G. \ch{CsSnI3} :
  {Semiconductor} or {Metal}? {High} {Electrical} {Conductivity} and {Strong}
  {Near}-{Infrared} {Photoluminescence} from a {Single} {Material}. {High}
  {Hole} {Mobility} and {Phase}-{Transitions}. \emph{J. Am. Chem. Soc.}
  \textbf{2012}, \emph{134}, 8579--8587\relax
\mciteBstWouldAddEndPuncttrue
\mciteSetBstMidEndSepPunct{\mcitedefaultmidpunct}
{\mcitedefaultendpunct}{\mcitedefaultseppunct}\relax
\EndOfBibitem
\bibitem[Gupta \latin{et~al.}(2016)Gupta, Bendikov, Hodes, and
  Cahen]{gupta_cssnbr_2016}
Gupta,~S.; Bendikov,~T.; Hodes,~G.; Cahen,~D. \ch{CsSnBr3} , {A} {Lead}-{Free}
  {Halide} {Perovskite} for {Long}-{Term} {Solar} {Cell} {Application}:
  {Insights} on {SnF} $_{\textrm{2}}$ {Addition}. \emph{ACS Energy Lett.}
  \textbf{2016}, \emph{1}, 1028--1033\relax
\mciteBstWouldAddEndPuncttrue
\mciteSetBstMidEndSepPunct{\mcitedefaultmidpunct}
{\mcitedefaultendpunct}{\mcitedefaultseppunct}\relax
\EndOfBibitem
\bibitem[Fabini \latin{et~al.}(2016)Fabini, Laurita, Bechtel, Stoumpos, Evans,
  Kontos, Raptis, Falaras, Van~der Ven, Kanatzidis, and
  Seshadri]{fabini_dynamic_2016}
Fabini,~D.~H.; Laurita,~G.; Bechtel,~J.~S.; Stoumpos,~C.~C.; Evans,~H.~A.;
  Kontos,~A.~G.; Raptis,~Y.~S.; Falaras,~P.; Van~der Ven,~A.;
  Kanatzidis,~M.~G.; Seshadri,~R. Dynamic {Stereochemical} {Activity} of the
  {Sn} $^{\textrm{2+}}$ {Lone} {Pair} in {Perovskite} \ch{CsSnBr3}. \emph{J.
  Am. Chem. Soc.} \textbf{2016}, \emph{138}, 11820--11832\relax
\mciteBstWouldAddEndPuncttrue
\mciteSetBstMidEndSepPunct{\mcitedefaultmidpunct}
{\mcitedefaultendpunct}{\mcitedefaultseppunct}\relax
\EndOfBibitem
\bibitem[Sercel \latin{et~al.}(2019)Sercel, Lyons, Wickramaratne, Vaxenburg,
  Bernstein, and Efros]{Sercel:2019je}
Sercel,~P.~C.; Lyons,~J.~L.; Wickramaratne,~D.; Vaxenburg,~R.; Bernstein,~N.;
  Efros,~Al.~L. {Exciton Fine Structure in Perovskite Nanocrystals}. \emph{Nano
  Lett.} \textbf{2019}, \emph{19}, 4068--4077\relax
\mciteBstWouldAddEndPuncttrue
\mciteSetBstMidEndSepPunct{\mcitedefaultmidpunct}
{\mcitedefaultendpunct}{\mcitedefaultseppunct}\relax
\EndOfBibitem
\bibitem[Ke \latin{et~al.}(2019)Ke, Stoumpos, and Kanatzidis]{ke_unleaded_2019}
Ke,~W.; Stoumpos,~C.~C.; Kanatzidis,~M.~G. “{Unleaded}” {Perovskites}:
  {Status} {Quo} and {Future} {Prospects} of {Tin}‐{Based} {Perovskite}
  {Solar} {Cells}. \emph{Adv. Mater.} \textbf{2019}, \emph{31}, 1803230\relax
\mciteBstWouldAddEndPuncttrue
\mciteSetBstMidEndSepPunct{\mcitedefaultmidpunct}
{\mcitedefaultendpunct}{\mcitedefaultseppunct}\relax
\EndOfBibitem
\bibitem[Monacelli and Marzari(2023)Monacelli, and
  Marzari]{monacelli_first-principles_2023}
Monacelli,~L.; Marzari,~N. First-{Principles} {Thermodynamics} of \ch{CsSnI3}.
  \emph{Chem. Mater.} \textbf{2023}, \emph{35}, 1702--1709\relax
\mciteBstWouldAddEndPuncttrue
\mciteSetBstMidEndSepPunct{\mcitedefaultmidpunct}
{\mcitedefaultendpunct}{\mcitedefaultseppunct}\relax
\EndOfBibitem
\bibitem[Fabini \latin{et~al.}(2020)Fabini, Seshadri, and
  Kanatzidis]{fabini_underappreciated_2020}
Fabini,~D.~H.; Seshadri,~R.; Kanatzidis,~M.~G. The underappreciated lone pair
  in halide perovskites underpins their unusual properties. \emph{MRS Bull.}
  \textbf{2020}, \emph{45}, 467--477\relax
\mciteBstWouldAddEndPuncttrue
\mciteSetBstMidEndSepPunct{\mcitedefaultmidpunct}
{\mcitedefaultendpunct}{\mcitedefaultseppunct}\relax
\EndOfBibitem
\bibitem[Lanigan-Atkins \latin{et~al.}(2021)Lanigan-Atkins, He, Krogstad,
  Pajerowski, Abernathy, Xu, Xu, Chung, Kanatzidis, Rosenkranz, Osborn, and
  Delaire]{lanigan-atkins_two-dimensional_2021}
Lanigan-Atkins,~T.; He,~X.; Krogstad,~M.~J.; Pajerowski,~D.~M.;
  Abernathy,~D.~L.; Xu,~G. N. M.~N.; Xu,~Z.; Chung,~D.-Y.; Kanatzidis,~M.~G.;
  Rosenkranz,~S.; Osborn,~R.; Delaire,~O. Two-dimensional overdamped
  fluctuations of the soft perovskite lattice in \ch{CsPbBr3}. \emph{Nat.
  Mater.} \textbf{2021}, \emph{20}, 977--983\relax
\mciteBstWouldAddEndPuncttrue
\mciteSetBstMidEndSepPunct{\mcitedefaultmidpunct}
{\mcitedefaultendpunct}{\mcitedefaultseppunct}\relax
\EndOfBibitem
\bibitem[Schilcher \latin{et~al.}(2021)Schilcher, Robinson, Abramovitch, Tan,
  Rappe, Reichman, and Egger]{schilcher_significance_2021}
Schilcher,~M.~J.; Robinson,~P.~J.; Abramovitch,~D.~J.; Tan,~L.~Z.;
  Rappe,~A.~M.; Reichman,~D.~R.; Egger,~D.~A. The {Significance} of {Polarons}
  and {Dynamic} {Disorder} in {Halide} {Perovskites}. \emph{ACS Energy Lett.}
  \textbf{2021}, \emph{6}, 2162--2173\relax
\mciteBstWouldAddEndPuncttrue
\mciteSetBstMidEndSepPunct{\mcitedefaultmidpunct}
{\mcitedefaultendpunct}{\mcitedefaultseppunct}\relax
\EndOfBibitem
\bibitem[Radha \latin{et~al.}(2018)Radha, Bhandari, and
  Lambrecht]{radha_distortion_2018}
Radha,~S.~K.; Bhandari,~C.; Lambrecht,~W. R.~L. Distortion modes in halide
  perovskites: {To} twist or to stretch, a matter of tolerance and lone pairs.
  \emph{Phys. Rev. Materials} \textbf{2018}, \emph{2}, 063605\relax
\mciteBstWouldAddEndPuncttrue
\mciteSetBstMidEndSepPunct{\mcitedefaultmidpunct}
{\mcitedefaultendpunct}{\mcitedefaultseppunct}\relax
\EndOfBibitem
\bibitem[Gao \latin{et~al.}(2021)Gao, Yadgarov, Sharma, Korobko, McCall,
  Fabini, Stoumpos, Kanatzidis, Rappe, and Yaffe]{gao_metal_2021}
Gao,~L.; Yadgarov,~L.; Sharma,~R.; Korobko,~R.; McCall,~K.~M.; Fabini,~D.~H.;
  Stoumpos,~C.~C.; Kanatzidis,~M.~G.; Rappe,~A.~M.; Yaffe,~O. Metal cation s
  lone-pairs increase octahedral tilting instabilities in halide perovskites.
  \emph{Mater. Adv.} \textbf{2021}, \emph{2}, 4610--4616\relax
\mciteBstWouldAddEndPuncttrue
\mciteSetBstMidEndSepPunct{\mcitedefaultmidpunct}
{\mcitedefaultendpunct}{\mcitedefaultseppunct}\relax
\EndOfBibitem
\bibitem[Fu \latin{et~al.}(2021)Fu, Jin, and Zhu]{fu_stereochemical_2021}
Fu,~Y.; Jin,~S.; Zhu,~X.-Y. Stereochemical expression of ns2 electron pairs in
  metal halide perovskites. \emph{Nat Rev Chem} \textbf{2021}, \emph{5},
  838--852\relax
\mciteBstWouldAddEndPuncttrue
\mciteSetBstMidEndSepPunct{\mcitedefaultmidpunct}
{\mcitedefaultendpunct}{\mcitedefaultseppunct}\relax
\EndOfBibitem
\bibitem[Marronnier \latin{et~al.}(2018)Marronnier, Roma, Boyer-Richard,
  Pedesseau, Jancu, Bonnassieux, Katan, Stoumpos, Kanatzidis, and
  Even]{marronnier_anharmonicity_2018}
Marronnier,~A.; Roma,~G.; Boyer-Richard,~S.; Pedesseau,~L.; Jancu,~J.-M.;
  Bonnassieux,~Y.; Katan,~C.; Stoumpos,~C.~C.; Kanatzidis,~M.~G.; Even,~J.
  Anharmonicity and {Disorder} in the {Black} {Phases} of {Cesium} {Lead}
  {Iodide} {Used} for {Stable} {Inorganic} {Perovskite} {Solar} {Cells}.
  \emph{ACS Nano} \textbf{2018}, \emph{12}, 3477--3486\relax
\mciteBstWouldAddEndPuncttrue
\mciteSetBstMidEndSepPunct{\mcitedefaultmidpunct}
{\mcitedefaultendpunct}{\mcitedefaultseppunct}\relax
\EndOfBibitem
\bibitem[Marronnier \latin{et~al.}(2019)Marronnier, Roma, Carignano,
  Bonnassieux, Katan, Even, Mosconi, and De~Angelis]{marronnier_influence_2019}
Marronnier,~A.; Roma,~G.; Carignano,~M.~A.; Bonnassieux,~Y.; Katan,~C.;
  Even,~J.; Mosconi,~E.; De~Angelis,~F. Influence of {Disorder} and
  {Anharmonic} {Fluctuations} on the {Dynamical} {Rashba} {Effect} in {Purely}
  {Inorganic} {Lead}-{Halide} {Perovskites}. \emph{J. Phys. Chem. C}
  \textbf{2019}, \emph{123}, 291--298\relax
\mciteBstWouldAddEndPuncttrue
\mciteSetBstMidEndSepPunct{\mcitedefaultmidpunct}
{\mcitedefaultendpunct}{\mcitedefaultseppunct}\relax
\EndOfBibitem
\bibitem[Steele \latin{et~al.}(2019)Steele, Puech, Monserrat, Wu, Yang,
  Kirchartz, Yuan, Fleury, Giovanni, Fron, Keshavarz, Debroye, Zhou, Sum,
  Walsh, Hofkens, and Roeffaers]{steele_role_2019}
Steele,~J.~A. \latin{et~al.}  Role of {Electron}–{Phonon} {Coupling} in the
  {Thermal} {Evolution} of {Bulk} {Rashba}-{Like} {Spin}-{Split} {Lead}
  {Halide} {Perovskites} {Exhibiting} {Dual}-{Band} {Photoluminescence}.
  \emph{ACS Energy Lett.} \textbf{2019}, \emph{4}, 2205--2212\relax
\mciteBstWouldAddEndPuncttrue
\mciteSetBstMidEndSepPunct{\mcitedefaultmidpunct}
{\mcitedefaultendpunct}{\mcitedefaultseppunct}\relax
\EndOfBibitem
\bibitem[Mohd~Yusoff \latin{et~al.}(2021)Mohd~Yusoff, Mahata, Vasilopoulou,
  Ullah, Hu, Jose Da~Silva, Kurt~Schneider, Gao, Ievlev, Liu, Ovchinnikova,
  De~Angelis, and Khaja~Nazeeruddin]{mohd_yusoff_observation_2021}
Mohd~Yusoff,~A. R.~B.; Mahata,~A.; Vasilopoulou,~M.; Ullah,~H.; Hu,~B.; Jose
  Da~Silva,~W.; Kurt~Schneider,~F.; Gao,~P.; Ievlev,~A.~V.; Liu,~Y.;
  Ovchinnikova,~O.~S.; De~Angelis,~F.; Khaja~Nazeeruddin,~M. Observation of
  large {Rashba} spin–orbit coupling at room temperature in compositionally
  engineered perovskite single crystals and application in high performance
  photodetectors. \emph{Materials Today} \textbf{2021}, \emph{46}, 18--27\relax
\mciteBstWouldAddEndPuncttrue
\mciteSetBstMidEndSepPunct{\mcitedefaultmidpunct}
{\mcitedefaultendpunct}{\mcitedefaultseppunct}\relax
\EndOfBibitem
\bibitem[Ambrosio \latin{et~al.}(2022)Ambrosio, De~Angelis, and
  Goñi]{ambrosio_ferroelectricferroelastic_2022}
Ambrosio,~F.; De~Angelis,~F.; Goñi,~A.~R. The {Ferroelectric}–{Ferroelastic}
  {Debate} about {Metal} {Halide} {Perovskites}. \emph{J. Phys. Chem. Lett.}
  \textbf{2022}, \emph{13}, 7731--7740\relax
\mciteBstWouldAddEndPuncttrue
\mciteSetBstMidEndSepPunct{\mcitedefaultmidpunct}
{\mcitedefaultendpunct}{\mcitedefaultseppunct}\relax
\EndOfBibitem
\bibitem[Sercel \latin{et~al.}(2019)Sercel, Lyons, Bernstein, and
  Efros]{Sercel:2019hs}
Sercel,~P.~C.; Lyons,~J.~L.; Bernstein,~N.; Efros,~Al.~L. {Quasicubic model for
  metal halide perovskite nanocrystals}. \emph{The Journal of Chemical Physics}
  \textbf{2019}, \emph{151}, 234106\relax
\mciteBstWouldAddEndPuncttrue
\mciteSetBstMidEndSepPunct{\mcitedefaultmidpunct}
{\mcitedefaultendpunct}{\mcitedefaultseppunct}\relax
\EndOfBibitem
\bibitem[Swift \latin{et~al.}(2021)Swift, Lyons, Efros, and
  Sercel]{swift_rashba_2021}
Swift,~M.~W.; Lyons,~J.~L.; Efros,~Al.~L.; Sercel,~P.~C. Rashba exciton in a
  {2D} perovskite quantum dot. \emph{Nanoscale} \textbf{2021}, \emph{13},
  16769--16780\relax
\mciteBstWouldAddEndPuncttrue
\mciteSetBstMidEndSepPunct{\mcitedefaultmidpunct}
{\mcitedefaultendpunct}{\mcitedefaultseppunct}\relax
\EndOfBibitem
\bibitem[Swift \latin{et~al.}(2023)Swift, Sercel, Efros, Lyons, and
  Norris]{swift_BGE_2023}
Swift,~M.~W.; Sercel,~P.~C.; Efros,~Al.~L.; Lyons,~J.~L.; Norris,~D.~J. Bright
  Excitons in Rashba Materials. \emph{(In preparation)} \textbf{2023}, \relax
\mciteBstWouldAddEndPunctfalse
\mciteSetBstMidEndSepPunct{\mcitedefaultmidpunct}
{}{\mcitedefaultseppunct}\relax
\EndOfBibitem
\bibitem[Isarov \latin{et~al.}(2017)Isarov, Tan, Bodnarchuk, Kovalenko, Rappe,
  and Lifshitz]{Isarov:2017do}
Isarov,~M.; Tan,~L.~Z.; Bodnarchuk,~M.~I.; Kovalenko,~M.~V.; Rappe,~A.~M.;
  Lifshitz,~E. {Rashba Effect in a Single Colloidal CsPbBr 3Perovskite
  Nanocrystal Detected by Magneto-Optical Measurements}. \emph{Nano Lett.}
  \textbf{2017}, \emph{17}, 5020--5026\relax
\mciteBstWouldAddEndPuncttrue
\mciteSetBstMidEndSepPunct{\mcitedefaultmidpunct}
{\mcitedefaultendpunct}{\mcitedefaultseppunct}\relax
\EndOfBibitem
\bibitem[Li \latin{et~al.}(2020)Li, Chen, Liu, Zhang, Chen, Wang, Yuan, and
  Feng]{li_evidence_2020}
Li,~X.; Chen,~S.; Liu,~P.-F.; Zhang,~Y.; Chen,~Y.; Wang,~H.-L.; Yuan,~H.;
  Feng,~S. Evidence for {Ferroelectricity} of {All}-{Inorganic} {Perovskite}
  {CsPbBr} $_{\textrm{3}}$ {Quantum} {Dots}. \emph{J. Am. Chem. Soc.}
  \textbf{2020}, \emph{142}, 3316--3320\relax
\mciteBstWouldAddEndPuncttrue
\mciteSetBstMidEndSepPunct{\mcitedefaultmidpunct}
{\mcitedefaultendpunct}{\mcitedefaultseppunct}\relax
\EndOfBibitem
\bibitem[Lv \latin{et~al.}(2021)Lv, Zhu, Tang, Lv, Zhang, Wang, Shu, and
  Xiao]{lv_probing_2021}
Lv,~B.; Zhu,~T.; Tang,~Y.; Lv,~Y.; Zhang,~C.; Wang,~X.; Shu,~D.; Xiao,~M.
  Probing {Permanent} {Dipole} {Moments} and {Removing} {Exciton} {Fine}
  {Structures} in {Single} {Perovskite} {Nanocrystals} by an {Electric}
  {Field}. \emph{Phys. Rev. Lett.} \textbf{2021}, \emph{126}, 197403\relax
\mciteBstWouldAddEndPuncttrue
\mciteSetBstMidEndSepPunct{\mcitedefaultmidpunct}
{\mcitedefaultendpunct}{\mcitedefaultseppunct}\relax
\EndOfBibitem
\bibitem[Liu \latin{et~al.}(2022)Liu, Gong, Geng, Feng, Manidaki, Deng,
  Stoumpos, Canepa, Xiao, Zhang, and Mao]{liu_hybrid_2022}
Liu,~Y.; Gong,~Y.; Geng,~S.; Feng,~M.; Manidaki,~D.; Deng,~Z.; Stoumpos,~C.~C.;
  Canepa,~P.; Xiao,~Z.; Zhang,~W.; Mao,~L. Hybrid {Germanium} {Bromide}
  {Perovskites} with {Tunable} {Second} {Harmonic} {Generation}. \emph{Angew
  Chem Int Ed} \textbf{2022}, \emph{61}\relax
\mciteBstWouldAddEndPuncttrue
\mciteSetBstMidEndSepPunct{\mcitedefaultmidpunct}
{\mcitedefaultendpunct}{\mcitedefaultseppunct}\relax
\EndOfBibitem
\bibitem[Stoumpos \latin{et~al.}(2015)Stoumpos, Frazer, Clark, Kim, Rhim,
  Freeman, Ketterson, Jang, and Kanatzidis]{stoumpos_hybrid_2015}
Stoumpos,~C.~C.; Frazer,~L.; Clark,~D.~J.; Kim,~Y.~S.; Rhim,~S.~H.;
  Freeman,~A.~J.; Ketterson,~J.~B.; Jang,~J.~I.; Kanatzidis,~M.~G. Hybrid
  {Germanium} {Iodide} {Perovskite} {Semiconductors}: {Active} {Lone} {Pairs},
  {Structural} {Distortions}, {Direct} and {Indirect} {Energy} {Gaps}, and
  {Strong} {Nonlinear} {Optical} {Properties}. \emph{J. Am. Chem. Soc.}
  \textbf{2015}, \emph{137}, 6804--6819\relax
\mciteBstWouldAddEndPuncttrue
\mciteSetBstMidEndSepPunct{\mcitedefaultmidpunct}
{\mcitedefaultendpunct}{\mcitedefaultseppunct}\relax
\EndOfBibitem
\bibitem[Perdew \latin{et~al.}(1996)Perdew, Burke, and Ernzerhof]{PBE}
Perdew,~J.~P.; Burke,~K.; Ernzerhof,~M. {Generalized gradient approximation
  made simple}. \emph{Phys. Rev. Lett} \textbf{1996}, \emph{77},
  3865--3868\relax
\mciteBstWouldAddEndPuncttrue
\mciteSetBstMidEndSepPunct{\mcitedefaultmidpunct}
{\mcitedefaultendpunct}{\mcitedefaultseppunct}\relax
\EndOfBibitem
\bibitem[Du(2015)]{Du:2015hs}
Du,~M.-H. {Density Functional Calculations of Native Defects in
  CH$_3$NH$_3$PbI$_3$: Effects of Spin{\textendash}Orbit Coupling and
  Self-Interaction Error}. \emph{J. Phys. Chem. Lett.} \textbf{2015}, \emph{6},
  1461--1466\relax
\mciteBstWouldAddEndPuncttrue
\mciteSetBstMidEndSepPunct{\mcitedefaultmidpunct}
{\mcitedefaultendpunct}{\mcitedefaultseppunct}\relax
\EndOfBibitem
\bibitem[Meggiolaro and De~Angelis(2018)Meggiolaro, and
  De~Angelis]{DeAngelis:2018gv}
Meggiolaro,~D.; De~Angelis,~F. {First-Principles Modeling of Defects in Lead
  Halide Perovskites: Best Practices and Open Issues}. \emph{ACS Energy Lett.}
  \textbf{2018}, \emph{3}, 2206--2222\relax
\mciteBstWouldAddEndPuncttrue
\mciteSetBstMidEndSepPunct{\mcitedefaultmidpunct}
{\mcitedefaultendpunct}{\mcitedefaultseppunct}\relax
\EndOfBibitem
\bibitem[Zhang \latin{et~al.}(2023)Zhang, Zhang, Turiansky, and Van
  De~Walle]{zhang_iodine_2023}
Zhang,~J.; Zhang,~X.; Turiansky,~M.~E.; Van De~Walle,~C.~G. Iodine {Vacancies}
  do not {Cause} {Nonradiative} {Recombination} in {Halide} {Perovskites}.
  \emph{PRX Energy} \textbf{2023}, \emph{2}, 013008\relax
\mciteBstWouldAddEndPuncttrue
\mciteSetBstMidEndSepPunct{\mcitedefaultmidpunct}
{\mcitedefaultendpunct}{\mcitedefaultseppunct}\relax
\EndOfBibitem
\bibitem[Lyons and Swift(2023)Lyons, and Swift]{lyons_trends_2023}
Lyons,~J.~L.; Swift,~M.~W. Trends for {Acceptor} {Dopants} in {Lead} {Halide}
  {Perovskites}. \emph{J. Phys. Chem. C} \textbf{2023}, \emph{127},
  12735--12740\relax
\mciteBstWouldAddEndPuncttrue
\mciteSetBstMidEndSepPunct{\mcitedefaultmidpunct}
{\mcitedefaultendpunct}{\mcitedefaultseppunct}\relax
\EndOfBibitem
\bibitem[Henderson \latin{et~al.}(2011)Henderson, Paier, and
  Scuseria]{henderson_accurate_2011}
Henderson,~T.~M.; Paier,~J.; Scuseria,~G.~E. Accurate treatment of solids with
  the {HSE} screened hybrid. \emph{phys. stat. sol. (b)} \textbf{2011},
  \emph{248}, 767--774\relax
\mciteBstWouldAddEndPuncttrue
\mciteSetBstMidEndSepPunct{\mcitedefaultmidpunct}
{\mcitedefaultendpunct}{\mcitedefaultseppunct}\relax
\EndOfBibitem
\bibitem[{H. Jonsson} \latin{et~al.}(1998){H. Jonsson}, {G. Mills}, and {K. W.
  Jacobsen}]{h_jonsson_nudged_1998}
{H. Jonsson},; {G. Mills},; {K. W. Jacobsen}, In \emph{Classical and {Quantum}
  {Dynamics} in {Condensed} {Phase} {Simulations}}; {B. J. Berne},, {G.
  Ciccotti},, {D. F. Coker},, Eds.; World Scientific, 1998\relax
\mciteBstWouldAddEndPuncttrue
\mciteSetBstMidEndSepPunct{\mcitedefaultmidpunct}
{\mcitedefaultendpunct}{\mcitedefaultseppunct}\relax
\EndOfBibitem
\bibitem[King-Smith and Vanderbilt(1993)King-Smith, and
  Vanderbilt]{King-Smith_Vanderbilt_Polarization_1993}
King-Smith,~R.~D.; Vanderbilt,~D. Theory of polarization of crystalline solids.
  \emph{Phys. Rev. B} \textbf{1993}, \emph{47}, 1651--1654\relax
\mciteBstWouldAddEndPuncttrue
\mciteSetBstMidEndSepPunct{\mcitedefaultmidpunct}
{\mcitedefaultendpunct}{\mcitedefaultseppunct}\relax
\EndOfBibitem
\bibitem[Stroppa \latin{et~al.}(2015)Stroppa, Quarti, De~Angelis, and
  Picozzi]{stroppa_ferroelectric_2015}
Stroppa,~A.; Quarti,~C.; De~Angelis,~F.; Picozzi,~S. Ferroelectric
  {Polarization} of {CH} $_{\textrm{3}}$ {NH} $_{\textrm{3}}$ {PbI}
  $_{\textrm{3}}$ : {A} {Detailed} {Study} {Based} on {Density} {Functional}
  {Theory} and {Symmetry} {Mode} {Analysis}. \emph{J. Phys. Chem. Lett.}
  \textbf{2015}, \emph{6}, 2223--2231\relax
\mciteBstWouldAddEndPuncttrue
\mciteSetBstMidEndSepPunct{\mcitedefaultmidpunct}
{\mcitedefaultendpunct}{\mcitedefaultseppunct}\relax
\EndOfBibitem
\bibitem[Wieder(1955)]{Wieder_1955}
Wieder,~H.~H. Electrical Behavior of Barium Titanate Single Crystals at Low
  Temperatures. \emph{Phys Rev} \textbf{1955}, \emph{99}, 1161--1165\relax
\mciteBstWouldAddEndPuncttrue
\mciteSetBstMidEndSepPunct{\mcitedefaultmidpunct}
{\mcitedefaultendpunct}{\mcitedefaultseppunct}\relax
\EndOfBibitem
\bibitem[Ducharme \latin{et~al.}(2000)Ducharme, Fridkin, Bune, Palto, Blinov,
  Petukhova, and Yudin]{ducharme_intrinsic_2000}
Ducharme,~S.; Fridkin,~V.~M.; Bune,~A.~V.; Palto,~S.~P.; Blinov,~L.~M.;
  Petukhova,~N.~N.; Yudin,~S.~G. Intrinsic {Ferroelectric} {Coercive} {Field}.
  \emph{Phys. Rev. Lett.} \textbf{2000}, \emph{84}, 175--178\relax
\mciteBstWouldAddEndPuncttrue
\mciteSetBstMidEndSepPunct{\mcitedefaultmidpunct}
{\mcitedefaultendpunct}{\mcitedefaultseppunct}\relax
\EndOfBibitem
\bibitem[Swarnkar \latin{et~al.}(2016)Swarnkar, Marshall, Sanehira,
  Chernomordik, Moore, Christians, Chakrabarti, and
  Luther]{swarnkar_quantum_2016}
Swarnkar,~A.; Marshall,~A.~R.; Sanehira,~E.~M.; Chernomordik,~B.~D.;
  Moore,~D.~T.; Christians,~J.~A.; Chakrabarti,~T.; Luther,~J.~M. Quantum
  dot–induced phase stabilization of $\alpha$-\ch{CsPbI3} perovskite for
  high-efficiency photovoltaics. \emph{Science} \textbf{2016}, \emph{354},
  92--95\relax
\mciteBstWouldAddEndPuncttrue
\mciteSetBstMidEndSepPunct{\mcitedefaultmidpunct}
{\mcitedefaultendpunct}{\mcitedefaultseppunct}\relax
\EndOfBibitem
\bibitem[Zhao \latin{et~al.}(2020)Zhao, Hazarika, Schelhas, Liu, Gaulding, Li,
  Zhang, Toney, Sercel, and Luther]{zhao_size-dependent_2020}
Zhao,~Q.; Hazarika,~A.; Schelhas,~L.~T.; Liu,~J.; Gaulding,~E.~A.; Li,~G.;
  Zhang,~M.; Toney,~M.~F.; Sercel,~P.~C.; Luther,~J.~M. Size-{Dependent}
  {Lattice} {Structure} and {Confinement} {Properties} in \ch{CsPbI3}
  {Perovskite} {Nanocrystals}: {Negative} {Surface} {Energy} for
  {Stabilization}. \emph{ACS Energy Lett.} \textbf{2020}, \emph{5},
  238--247\relax
\mciteBstWouldAddEndPuncttrue
\mciteSetBstMidEndSepPunct{\mcitedefaultmidpunct}
{\mcitedefaultendpunct}{\mcitedefaultseppunct}\relax
\EndOfBibitem
\bibitem[Jia \latin{et~al.}(2022)Jia, Chen, Zhuang, Hua, and
  Zhang]{jia_inhibiting_2022}
Jia,~D.; Chen,~J.; Zhuang,~R.; Hua,~Y.; Zhang,~X. Inhibiting lattice distortion
  of \ch{CsPbI3} perovskite quantum dots for solar cells with efficiency over
  16.6\%. \emph{Energy Environ. Sci.} \textbf{2022}, \emph{15},
  4201--4212\relax
\mciteBstWouldAddEndPuncttrue
\mciteSetBstMidEndSepPunct{\mcitedefaultmidpunct}
{\mcitedefaultendpunct}{\mcitedefaultseppunct}\relax
\EndOfBibitem
\bibitem[Mori and Saito(1986)Mori, and Saito]{mori_x-ray_1986}
Mori,~M.; Saito,~H. An {X}-ray study of successive phase transitions in
  {CsSnBr} $_{\textrm{3}}$. \emph{J. Phys. C: Solid State Phys.} \textbf{1986},
  \emph{19}, 2391--2401\relax
\mciteBstWouldAddEndPuncttrue
\mciteSetBstMidEndSepPunct{\mcitedefaultmidpunct}
{\mcitedefaultendpunct}{\mcitedefaultseppunct}\relax
\EndOfBibitem
\bibitem[Kresse and Furthm{\"u}ller(1996)Kresse, and Furthm{\"u}ller]{VASP}
Kresse,~G.; Furthm{\"u}ller,~J. {Efficient iterative schemes for ab initio
  total-energy calculations using a plane-wave basis set}. \emph{Phys. Rev. B}
  \textbf{1996}, \emph{54}, 11169--11186\relax
\mciteBstWouldAddEndPuncttrue
\mciteSetBstMidEndSepPunct{\mcitedefaultmidpunct}
{\mcitedefaultendpunct}{\mcitedefaultseppunct}\relax
\EndOfBibitem
\bibitem[Bl{\"o}chl(1994)]{PAW}
Bl{\"o}chl,~P.~E. {Projector Augmented-Wave Method}. \emph{Phys. Rev. B}
  \textbf{1994}, \emph{50}, 17953--17979\relax
\mciteBstWouldAddEndPuncttrue
\mciteSetBstMidEndSepPunct{\mcitedefaultmidpunct}
{\mcitedefaultendpunct}{\mcitedefaultseppunct}\relax
\EndOfBibitem
\bibitem[Heyd \latin{et~al.}(2003)Heyd, Scuseria, and Ernzerhof]{HSE}
Heyd,~J.; Scuseria,~G.~E.; Ernzerhof,~M. {Hybrid functionals based on a
  screened Coulomb potential}. \emph{J. Chem. Phys.} \textbf{2003}, \emph{118},
  8207--8215\relax
\mciteBstWouldAddEndPuncttrue
\mciteSetBstMidEndSepPunct{\mcitedefaultmidpunct}
{\mcitedefaultendpunct}{\mcitedefaultseppunct}\relax
\EndOfBibitem
\bibitem[Momma and Izumi(2011)Momma, and Izumi]{momma_vesta_2011}
Momma,~K.; Izumi,~F. \textit{{VESTA} 3} for three-dimensional visualization of
  crystal, volumetric and morphology data. \emph{J Appl Crystallogr}
  \textbf{2011}, \emph{44}, 1272--1276\relax
\mciteBstWouldAddEndPuncttrue
\mciteSetBstMidEndSepPunct{\mcitedefaultmidpunct}
{\mcitedefaultendpunct}{\mcitedefaultseppunct}\relax
\EndOfBibitem
\bibitem[Togo and Tanaka(2018)Togo, and Tanaka]{togo_textttspglib_2018}
Togo,~A.; Tanaka,~I. $\texttt{Spglib}$: a software library for crystal symmetry
  search. 2018; arXiv:1808.01590 [cond-mat]\relax
\mciteBstWouldAddEndPuncttrue
\mciteSetBstMidEndSepPunct{\mcitedefaultmidpunct}
{\mcitedefaultendpunct}{\mcitedefaultseppunct}\relax
\EndOfBibitem
\bibitem[Ong \latin{et~al.}(2013)Ong, Richards, Jain, Hautier, Kocher, Cholia,
  Gunter, Chevrier, Persson, and Ceder]{Ong2013}
Ong,~S.~P.; Richards,~W.~D.; Jain,~A.; Hautier,~G.; Kocher,~M.; Cholia,~S.;
  Gunter,~D.; Chevrier,~V.~L.; Persson,~K.~A.; Ceder,~G. Python Materials
  Genomics (pymatgen): A robust, open-source python library for materials
  analysis. \emph{Comput. Mater. Sci.} \textbf{2013}, \emph{68}, 314--319\relax
\mciteBstWouldAddEndPuncttrue
\mciteSetBstMidEndSepPunct{\mcitedefaultmidpunct}
{\mcitedefaultendpunct}{\mcitedefaultseppunct}\relax
\EndOfBibitem
\bibitem[Stoumpos \latin{et~al.}(2013)Stoumpos, Malliakas, Peters, Liu,
  Sebastian, Im, Chasapis, Wibowo, Chung, Freeman, Wessels, and
  Kanatzidis]{Stoumpos:2013fz}
Stoumpos,~C.~C.; Malliakas,~C.~D.; Peters,~J.~A.; Liu,~Z.; Sebastian,~M.;
  Im,~J.; Chasapis,~T.~C.; Wibowo,~A.~C.; Chung,~D.~Y.; Freeman,~A.~J.;
  Wessels,~B.~W.; Kanatzidis,~M.~G. {Crystal Growth of the Perovskite
  Semiconductor CsPbBr$_3$: A New Material for High-Energy Radiation
  Detection}. \emph{Cryst. Growth Des.} \textbf{2013}, \emph{13},
  2722--2727\relax
\mciteBstWouldAddEndPuncttrue
\mciteSetBstMidEndSepPunct{\mcitedefaultmidpunct}
{\mcitedefaultendpunct}{\mcitedefaultseppunct}\relax
\EndOfBibitem
\bibitem[Zhao \latin{et~al.}(2019)Zhao, Xu, Zhou, Li, Zhang, Xia, Liu, Dai, and
  Yao]{zhao_preparation_2019}
Zhao,~H.; Xu,~J.; Zhou,~S.; Li,~Z.; Zhang,~B.; Xia,~X.; Liu,~X.; Dai,~S.;
  Yao,~J. Preparation of {Tortuous} {3D} $\gamma$‐\ch{CsPbI3} {Films} at
  {Low} {Temperature} by {CaI} $_{\textrm{2}}$ as {Dopant} for {Highly}
  {Efficient} {Perovskite} {Solar} {Cells}. \emph{Adv. Funct. Mater.}
  \textbf{2019}, 1808986\relax
\mciteBstWouldAddEndPuncttrue
\mciteSetBstMidEndSepPunct{\mcitedefaultmidpunct}
{\mcitedefaultendpunct}{\mcitedefaultseppunct}\relax
\EndOfBibitem
\end{mcitethebibliography}
\end{document}